\documentclass[aps,prl,superscriptaddress,twocolumn]{revtex4-1}
\usepackage{units}
\usepackage{amsmath}
\usepackage{amssymb}
\usepackage{graphicx}
\usepackage{bm,color,ulem,microtype}

\newcommand{\gper}{\gamma_\perp}
\newcommand{\gpar}{\gamma_\parallel}
\newcommand{\be}{\begin{equation}}
\newcommand{\ee}{\end{equation}}
\newcommand{\bea}{\begin{eqnarray}}
\newcommand{\eea}{\end{eqnarray}}

\newcommand{\eq}[1]{(\ref{#1})}
\newcommand{\Eq}[1]{Equation~(\ref{#1})}
\newcommand{\Eqs}[1]{Equations~(\ref{#1})}

\newcommand{\Fig}[1]{Fig.~\ref{#1}}

\newcommand{\beshp}[2]{\mbox{\it H}^{+}_{#1}(#2)}    

\newcommand{\re}[1]{\text{Re}[#1]}

\newcommand{\si}{{SI}}

\begin{document}
\title{\textsf{Enhancement of Laser Power Efficiency by Control of Spatial Hole Burning Interactions}}
\author{\textsf{Li Ge}}
\affiliation{\textsf{Department of Electrical Engineering, Princeton University, Princeton, New Jersey 08544, USA}}
\affiliation{\textsf{Department of Engineering Science and Physics, College of Staten Island, CUNY, Staten Island, NY 10314, USA}}
\affiliation{\textsf{The Graduate Center, CUNY, New York, NY 10016, USA}}
\author{\textsf{Omer Malik}}
\affiliation{\textsf{Department of Electrical Engineering, Princeton University, Princeton, New Jersey 08544, USA}}
\author{\textsf{Hakan E. T\"ureci}}
\affiliation{\textsf{Department of Electrical Engineering, Princeton University, Princeton, New Jersey 08544, USA}}

\begin{abstract}
\textsf{The laser is an out-of-equilibrium nonlinear wave system where the interplay of the cavity geometry and nonlinear wave interactions, mediated by the gain medium, determines the self-organized oscillation frequencies and the associated spatial field patterns. In the steady state, a constant energy flux flows through the laser from the pump to the far field, with the ratio of the total output power to the input power determining the power-efficiency. While nonlinear wave interactions have been modelled and well understood since the early days of laser theory, their impact on the power-efficiency of a laser system is poorly understood. Here, we show that spatial hole burning interactions generally decrease the power efficiency. We then demonstrate how spatial hole burning interactions can be controlled by a spatially tailored pump profile, thereby boosting the power-efficiency, in some cases by orders of magnitude.}
\end{abstract}

\maketitle

Power-efficiency of lasers is a key property to be maximized to reduce energy requirements for on-chip applications, facilitate thermal management and pack lasers into smaller volumes. Earlier work on solid-state lasers addressed the quantum design of the gain medium and the optimization of carrier transport, demonstrating strong improvements in power-efficiency \cite{DIODE, razeghi}. Far less systematic work has addressed the fundamental factors limiting the power efficiency of lasers and effective methods to overcome these, taking into account specific resonator properties. In particular, the impact of spatial hole burning interactions is poorly understood \cite{e1}. A case in point are the findings of Ref.~\cite{gmachl_science98}, where the output power of microcavity quantum cascade lasers was found to increase exponentially with boundary deformation, resulting in a power enhancement over two orders of magnitude with respect to identical lasers of circular cross-section. After more than a decade of theoretical development, the fundamental factors behind this dramatic increase in power efficiency remain unknown. The goal of this paper is to provide a theoretical analysis of the mechanisms at work that enable such dramatic improvements in laser power efficiency.

\begin{figure}[b]
\includegraphics[clip,width=1.0\linewidth]{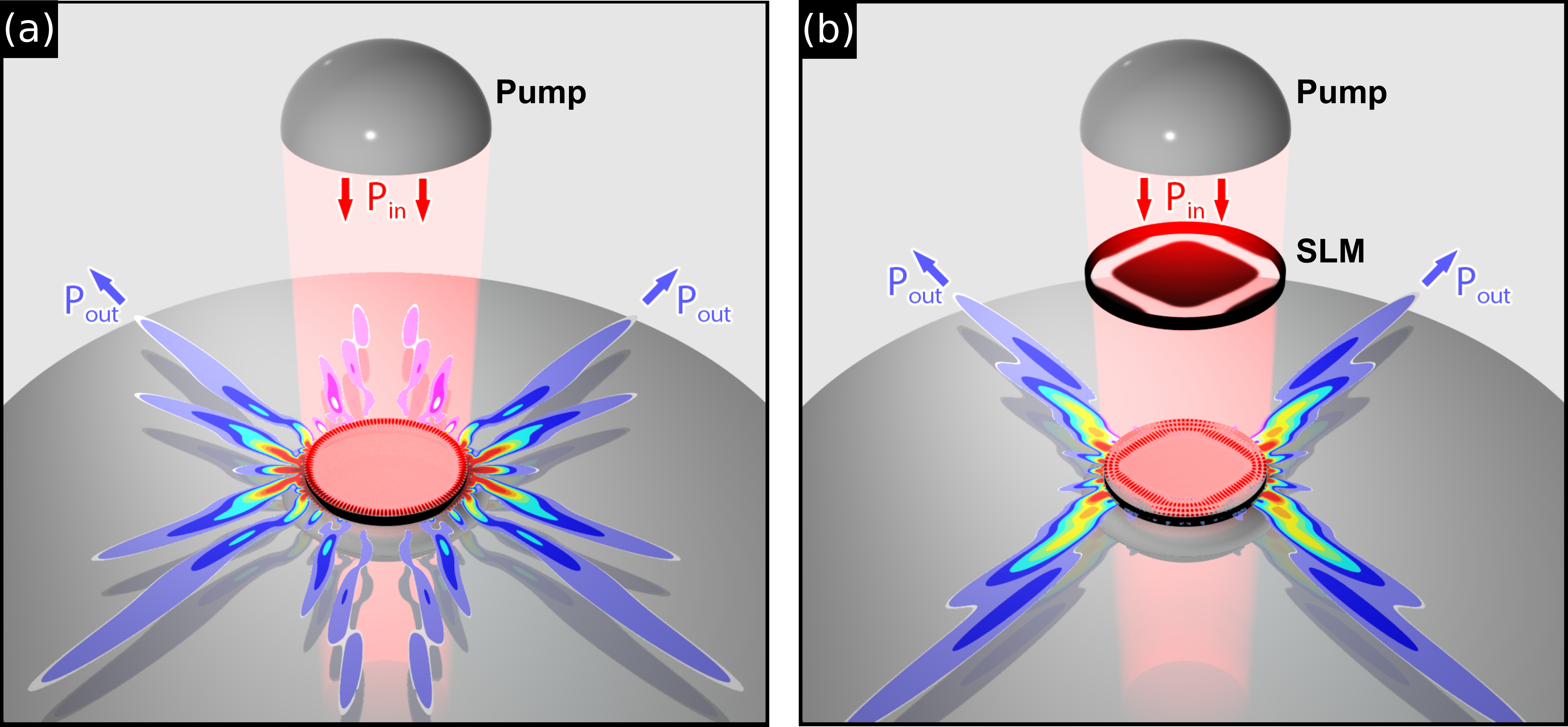}
\caption{A comparison of different optical pumping schemes and their corresponding emission patterns. (a) A high-Q whispering-gallery mode is excited by the uniform pump applied and in (b) a spatial-light modulator (SLM) is used to shape the beam and select a diamond-shaped mode. Selective pumping can also be achieved by a special lens (see Ref.~\cite{chern}) that focuses the total incoming power to a reduced total area or by patterned contacts in the case of current-injection lasers, which are the two mechanisms discussed in the text.
} \label{fig:povray}
\end{figure}
Typically, the pump power in a microlaser is deposited uniformly across the entire cavity area [\Fig{fig:povray}(a)] and lasing naturally occurs in cavity modes with the longest cavity lifetimes. Here we show the existence of a maximally power-efficient ''optimally out-coupled mode" that may never turn on in a uniformly pumped cavity due to spatial-hole burning interactions. We further show that a non-uniform pump distribution designed to selectively pump the optimally out-coupled mode can lead to strong power enhancements.

Spatially non-uniform pump distributions can be realized by fabricating patterned contacts \cite{kneissl}, by spatially non-uniform doping in the case of current injection lasers \cite{gerace,vuckovic}, or by using spatial light modulators or special lenses for optically pumped lasers \cite{Rex_IEEE01,chern}  (see \Fig{fig:povray}(b)). We discuss the optimal spatial profile of the pump for a given resonator to maximize the power efficiency, assuming a fixed internal quantum efficiency for the gain medium. We demonstrate the proof of principle for a whispering gallery resonator where a remarkable power enhancement factor of 185 can be achieved. The power enhancements we quote below are the relative power enhancements of the total multi-mode laser power (summed over all modes) for the selective pumping (\Fig{fig:povray}(b)) versus the uniform pumping (\Fig{fig:povray}(a)) of the identical resonator with the same gain volume.

\vspace{8pt}
\noindent\textbf{Multi-mode lasing in open systems} -- Our analysis is based on a recent reformulation of the semiclassical laser equations, the Steady-state Ab-initio Laser Theory (SALT) \cite{TS,Science,SPASALT}, which allows us to describe lasing characteristics that are independent of the specific details of the gain medium (details in Section 1 of Supplementary Information (SI); Ref.~\cite{SI}).

SALT finds the steady-state solution of the semiclassical laser equations in the general multi-periodic form
\begin{equation}
    E^+(\vec{r},t) = \sum_{\mu=1}^N \Psi_\mu(\vec{r})\,e^{-i\Omega_\mu t},
    P^+(\vec{r},t) = \sum_{\mu=1}^N p_\mu(\vec{r})\,e^{-i\Omega_\mu t},
    \nonumber
\end{equation}
where $N$ is the number of lasing modes at a given pump power, and $E^+(\vec{r},t)$ and $P^+(\vec{r},t)$ are the positive frequency components of the electric field and polarization density. The non-linear lasing modes $\Psi_\mu$ and the laser frequencies $\Omega_\mu$ can be obtained from the following set of coupled Helmholtz equations for a homogeneously broadened gain medium:
\begin{equation}
\left[ \nabla^2 + (\epsilon_c(\vec{r}) + \epsilon_g(\vec{r}))\Omega_\mu^2 \right]\Psi_\mu(\vec{r}) = 0, ~ ~ \vec{r} \in \text{cavity} \label{eq:SALT1}
\end{equation}
with an effective dielectric function consisting of the passive contribution $\epsilon_c(\vec{r})=n_c^2(\vec{r})$ and an active contribution $\epsilon_g(\vec{r})$ from the gain medium
\begin{equation}
\epsilon_g(\vec{r}) = \frac{\gper}{\Omega_\mu - \omega_a + i\gper}\frac{D_0 (\vec{r})}{1+\sum_{\nu=1}^N\Gamma_\nu|\Psi_\nu(\vec{r})|^2}. \label{eq:epsg}
\end{equation}
These equations are to be interpreted as a set of non-linear eigenvalue equations to be solved for the lasing frequencies $\Omega_\mu$ and the non-linear lasing modes $\Psi_\mu$ for a given pump distribution $D_0(\vec{r})$, capturing directly the steady state behavior. The complex linear index of refraction $n_c(\vec{r})$ defines the cavity and $\epsilon_g(\vec{r})$ contains the non-linear spatial hole-burning interactions exactly, to infinite order in the intensity of the lasing modes. We have taken the speed of light in vacuum $c=1$, $\omega_a$ is the atomic transition frequency, $\gpar$ and $\gamma_{\perp}$ are the inversion and polarization relaxation rates, $g$ is the transition dipole matrix element between the lasing levels, and $\Gamma_\mu \equiv \gper^2/[\gper^2 + (\Omega_\mu-\omega_a)^2]$ is the Lorentzian gain curve evaluated at the lasing frequency of the lasing mode $\mu$. The field $\Psi_{\mu}(\vec{r})$ is dimensionless and measured in units of $E_0 = \hbar\sqrt{\gpar\gper}/2g$. The pump is given by $D_0(\vec{r}) = D_0f(\vec{r})$, where $D_0$  measures the total power injected in terms of the spatially averaged inversion density and is kept the same when comparing the output power between uniformly and selectively pumped cavities (\si~Section 1.1). $f(\vec{r})$ is the spatial profile of the pump.

\Eqs{eq:SALT1} describe the radiation field both inside and outside the cavity and their direct numerical solution on a grid generally scales with the size of the entire scattering domain. The original semiclassical laser equations (\si~Eqs. (S1-S3)) as well as the steady-state equations \eq{eq:SALT1} can however be solved also by projecting them onto a suitable linear basis that scales only with the size of the cavity. This is what is done here, accounting for openness {\it exactly} by expanding each lasing mode $\Psi_\mu(\vec{r})$ further into a non-Hermitian basis set $\{\phi_m(\vec{r})\}$, the constant-flux (CF) basis \cite{TS},
\begin{equation}
\Psi_\mu(\vec{r}) = \sum_m a_m^\mu\phi_m(\vec{r};\Omega_\mu).
\label{eq:sumcf}
\end{equation}
This basis as well as the associated complex-valued eigenfrequencies $\omega_\mu = \nu_\mu - i\kappa_\mu$ are parametrically dependent on $\Omega_\mu$. We note that the CF basis is a complete set of states to expand an arbitrary field that oscillates at frequency $\Omega_\mu$ inside an open cavity, and is different from  quasi-bound modes (poles of the scattering matrix) \cite{ching_rmp98} because they carry a constant flux to the farfield. In the following sections, we further develop SALT projected onto the CF basis, to determine and excite the optimally out-coupled mode.

\vspace{8pt}
\noindent\textbf{Optimally out-coupled mode} -- The optimally out-coupled mode of a given resonator is the mode that provides the largest slope-efficiency for the total optical power in the farfield. The farfield power radiated by a lasing mode $\mu$ at a given pump power $D_0$ can be expressed as $P_\text{out} = C_\mu|a_\mu(D_0)|^2$, given that the lasing mode can be approximated by a single CF state, typically an excellent approximation for high-Q cavities ({\si} Section 1.1). We call $C_\mu$ the {\it out-coupling efficiency} for mode $\mu$ and find by calculation that it is inversely proportional to the {\it out-coupling Q-factor} $Q_\mu \equiv \nu_\mu/2\kappa_\mu$ ({\si} Section 1.2 and Fig.~S1), defined using the complex-valued resonances $\bar\omega_\mu\approx \nu_\mu - i\kappa_\mu$ in the absence of absorption.

For a microcavity of arbitrary shape and uniform index of refraction $n_c = \sqrt{\epsilon_c} \equiv n_0 + in_1$ ($n_0 \gg n_1 > 0$), the threshold of mode $\mu$ is given by
\begin{equation}
D_{th}^{\mu} \approx {n_0^2}(Q_\mu^{-1}+Q_\text{abs}^{-1}) \equiv D^\mu_\text{out} + D_\text{abs}, \label{eq:TH_absorp}
\end{equation}
where $Q_\text{abs} \equiv n_0/2n_1$ is the {\it absorptive $Q$-factor} [see \Fig{fig:Qopt}(a)] neglecting spatial hole-burning interactions.

Now considering $P_\text{out}$ and $D^\mu_{th}$ together, we note that while a large $Q_\mu$ implies a low threshold (and hence a large $|a_\mu|^2$) it also gives rise to a low out-coupling efficiency $C_\mu$. The total output power $P_\text{out}$ is maximized at an optimal value of $Q_\mu$, the {\it optimal $Q$-factor} ({\si} Section 1.2)
\begin{equation}
Q_\text{opt} = \frac{Q_\text{abs}}{\sqrt{\beta} - 1},\label{eq:Qopt}
\end{equation}
where $\beta\equiv D_0/D_\text{abs}$ is the pump strength in terms of the minimum threshold given by \Eq{eq:TH_absorp}.
Equation~({\ref{eq:Qopt}) shows that the optimal $Q$-factor is close to the absorptive $Q$-value. This rule of thumb is well-known and can be rigorously derived for a Fabry-Perot cavity \cite{siegman_book, milonni_book,Rigrod}. Similar out coupling criteria in the absence of self-saturation have also been used for fiber-coupled microcavity lasers \cite{Yariv,Min,Srinivasan}. Here we presented the rigorous foundations for this rule and generalized this condition to an arbitrarily complex open cavity, taking into account self-saturation effects -- the latter manifested in the pump dependence (viz. $\beta$) of the optimal out-coupling condition [see \Fig{fig:Qopt}(b)].

\begin{figure}[t]
\includegraphics[clip,width=1.0\linewidth]{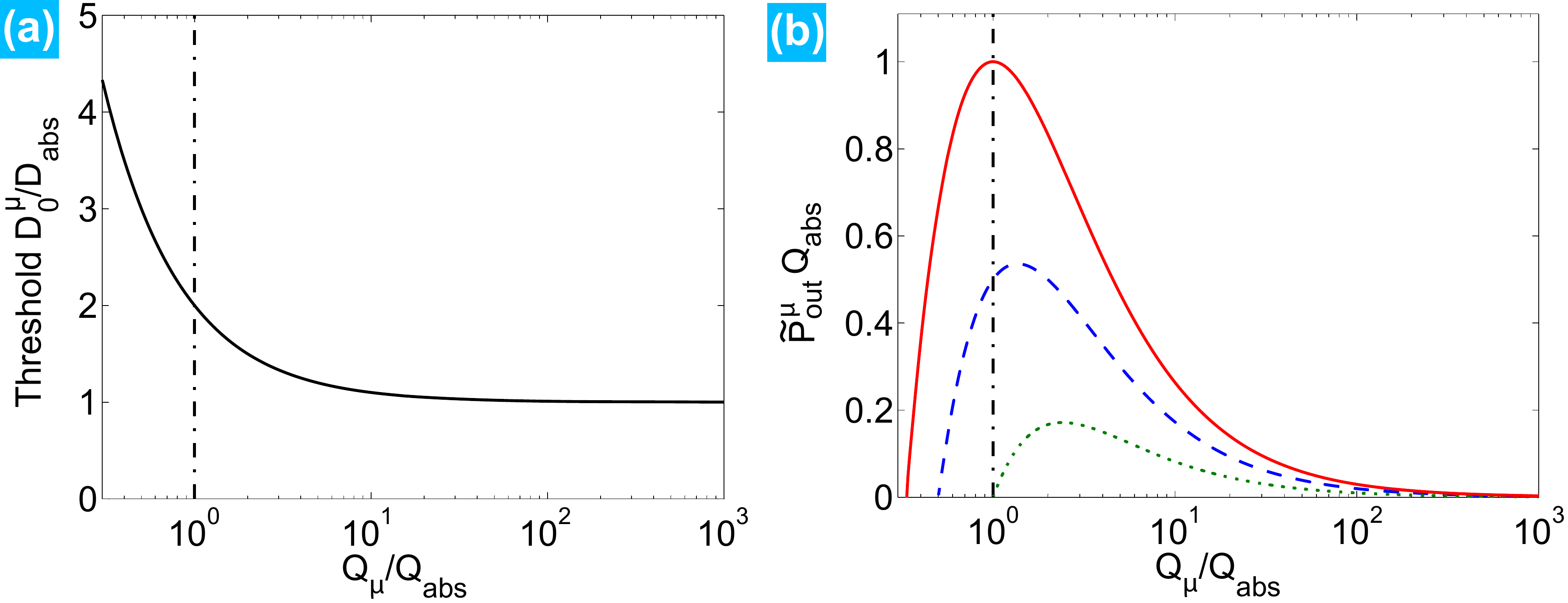}
\caption{Characteristics of the optimally out-coupled mode. The threshold (a) is low while the output intensity (b) peaks. (a) Thresholds as a function of the outcoupling $Q$-factor for uniform pumping (modal interactions neglected). The vertical dash-dotted line marks the absorptive $Q$-value. (b) Normalized output intensity curves shown for three values of pump strength: $\beta=2$ (dotted), 3 (dashed), and 4 (solid). $\tilde{P}_\text{out}^\mu$ is the output power given by Equation S35. For this calculation $\chi_\mu$ and $\Gamma_\mu$ are taken to be 1.
}\label{fig:Qopt}
\end{figure}

\vspace{8pt}
\noindent\textbf{Mode discrimination by spatially selective pumping} -- Microlasers typically operate in the multimode regime where spatial hole-burning interactions with lower-threshold modes can entirely suppress the optimally out-coupled mode, identified above in the absence of these interactions. Here we discuss how the optimally out-coupled mode can be favored to lase and the rest of the modes suppressed by spatially tailoring the pump profile.

To this end, we resort to finding an ideal spatial pump profile that overlaps strongly with the optimally out-coupled mode. Spatially nonuniform pumping schemes have previously been used to control various lasing characteristics, including emission directionality, thresholds, and frequencies in microdisks \cite{Rex_IEEE01}, spirals \cite{chern,Kwon_OL06}, asymmetric resonant cavities (ARC) \cite{fukushima_optlett02, Shinohara_PRL10}, and random lasers \cite{Xu2006, Ge_thesis,Rotter2012,RL2012,gainState,gainState2}, mostly in the single-mode regime and generally in a rather ad-hoc manner. To our knowledge, the present work is the first one to propose it as a scheme to systematically boost laser power in the multi-mode regime.

Consider a normalized spatial pump profile $f(\vec{r})$ in the plane of the microcavity. In Section 1.3 of the \si, we show that the threshold of mode $\mu$, in terms of the total pump power delivered to the entire cavity area, is modified from $D^\mu_{th}$ to approximately $D^\mu_{th}/\re{f_\mu}$. Here the complex-valued {\it pump overlap factor} is defined by
\begin{equation}
f_\mu\equiv \int_\text{cavity}f(\vec{r})\phi_\mu^2(\vec{r}) d\vec{r},
\end{equation}
Once the optimally outcoupled mode has been identified for a given cavity, we introduce a spatial pump profile that leads to a large overlap factor for this target mode, giving rise to a {\it pump-focusing effect} that reduces its threshold. For the higher-$Q$ modes that have lower thresholds with a uniform pump profile, the overlap factors can in principle be made smaller than unity, leading to increased thresholds inhibiting unwanted modes from lasing.

\vspace{8pt}
\noindent\textbf{Case study: the whispering gallery microlaser} -- For a proof of principle demonstration of our approach, we consider selectively pumping a microdisk laser, treated as a two-dimensional system using the effective index approach. This is a case where analytical formulas can be obtained for many of the factors discussed in previous sections. We find here the full non-linear steady-state solutions of the semiclassical laser equations expressed in \Eq{eq:SALT1} ({\si} Section 1.1).

Microdisk lasers utilize total internal reflection to trap light inside for a long time, and the field distribution of each high-$Q$ whispering-gallery mode changes little with non-linearity. In this case, a lasing mode can be represented by a single CF state,
characterized by its angular and radial quantum numbers $\mu = (m, \eta)$:
\begin{equation}
        \phi_\mu(r,\phi) = \begin{cases} \frac{1}{\zeta_\mu}J_{m}(n_c \omega_\mu r)e^{im \phi},  & r < R, \\
        \frac{1}{\zeta_\mu}\frac{J_{m}(n_c \omega_\mu R)}{H^+_{m}(\Omega_\mu R)}\,\beshp{m}{\Omega_\mu r} e^{im \phi}, & r > R,\end{cases}\label{eq:CF2D}
\end{equation}
where $J_m,H_m^+$ are Bessel and Hankel functions of the first kind and $\zeta_\mu = [2\pi\int_{\text{cavity}}J_m^2(n_c \omega_\mu r)~rdr]^{1/2}$ is a normalization factor. The real-valued lasing frequency $\Omega_\mu$ and the complex-valued CF frequency $\omega_\mu$ are related by a generalized frequency pulling expression ({\si} Eq.~S24). The exact expression for the out-coupling efficiency (as defined in the previous section) associated with each mode is
\begin{gather}
C_\mu = G\frac{\omega_a}{\Omega_\mu}\left|\frac{1}{\zeta_\mu}\frac{J_{m}(n_c\omega_\mu R)}{\beshp{m}{\Omega_\mu R}}\right|^2,
\end{gather}
where $G$ is a geometry-dependent factor defined in Eq.~S34.

\begin{figure*}[t]
\includegraphics[clip,width=.8\linewidth]{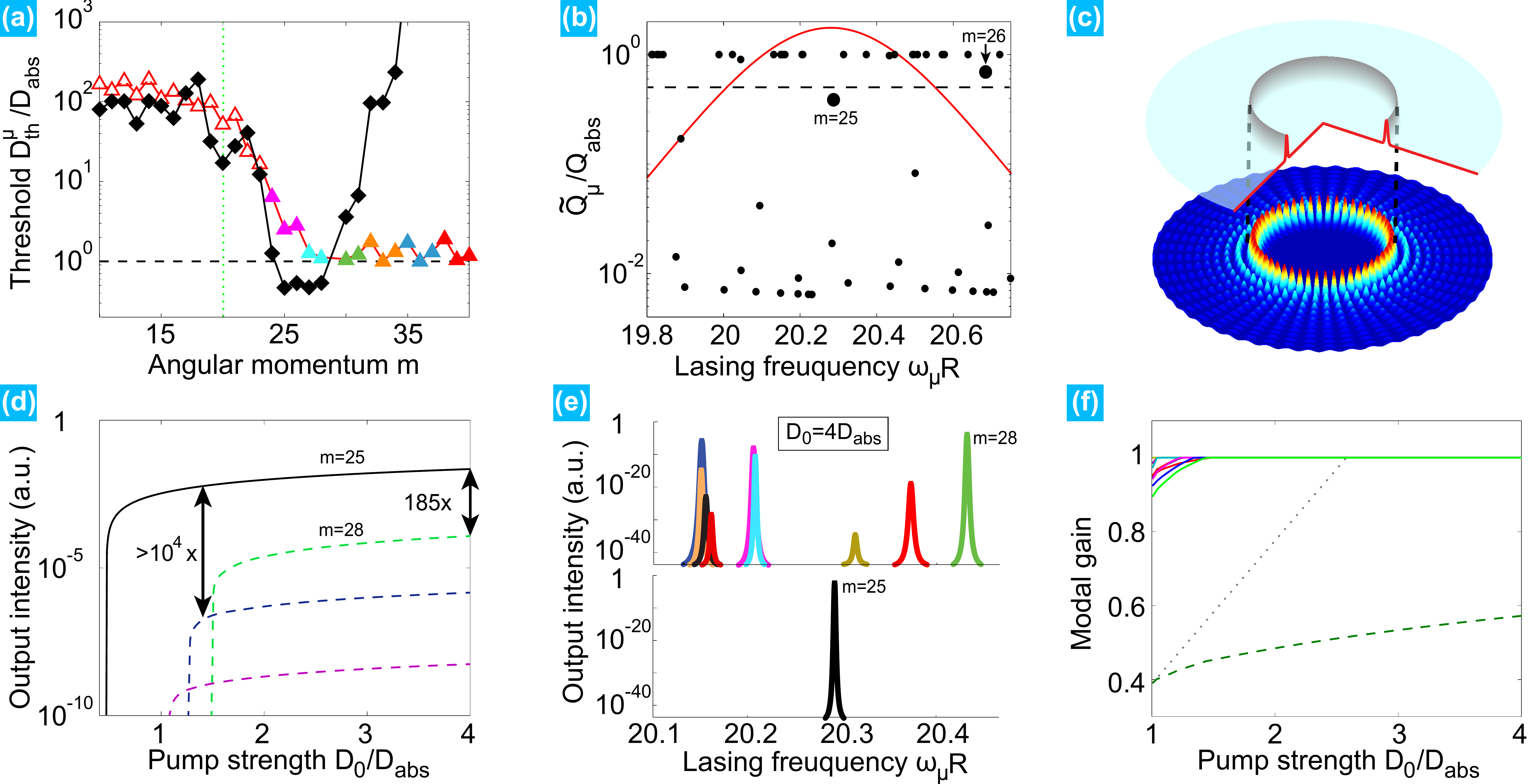}
\caption{Characteristics of a laser with $n_c=3.3+10^{-4}i$, $\omega_aR=20.28$, $\gper R=0.5$ (a) Single-mode thresholds versus $m$ for uniform (triangles) versus selective (diamond) pumping. Color-grouped triangles indicate modes with the same radial quantum number $\eta$, and vertical line marks the mode with $\chi_\mu = \chi_c$. (b) Total $Q$-factors versus mode frequency shown under gain curve; dashed line marks $Q_{opt}$. (c) Intensity distribution of mode $m=25$ and the corresponding pump profile. (d) Output intensity versus pump for the selectively pumped laser (solid line) and the three strongest modes of the uniformly pumped laser (dashed lines). (e) Spectrum at $D_0=4D_{abs}$ with uniform pumping (top) and selective pumping (bottom); linewidths are schematic. (f) Solid lines show modal gain for the uniformly pumped laser (see \cite{Science,SPASALT}). Dotted line indicates that $m=25$ turns on at $D_0=2.58D_{abs}$ in the absence of modal interactions (dotted) and well outside the shown pump range if modal interactions are included (dashed).}\label{fig:disk}
\end{figure*}

Under uniform pumping, the threshold of mode $\mu$ in the single-mode operation reduces exponentially as the incidence angle $\chi_\mu \approx \sin^{-1}(m/n_0\Omega_\mu R)$ becomes larger than the critical angle $\chi_c = \sin^{-1}(1/n_0)$ [\Fig{fig:disk}(a)], until it reaches the minimum threshold $D_\text{abs}$. The small fluctuation of the thresholds near the asymptote is due to the finite width of the gain; modes closer to the gain center $\omega_a$ [\Fig{fig:disk}(b)] have slightly lower thresholds, which is not considered in the approximate expression~(\ref{eq:TH_absorp}).

We take the cavity refractive index $n_c=3.3+10^{-4}i$, which leads to an absorptive $Q$-factor $Q_\text{abs}=1.65\times10^4$. Since the out-coupling $Q$-factors can be many orders of magnitude larger than $Q_\text{abs}$, it is more convenient to compare the total $Q$-factor,
\begin{equation}
\tilde{Q}_\mu \equiv \frac{Q_\mu Q_\text{abs}}{Q_\mu + Q_\text{abs}},
\end{equation}
with $Q_\text{abs}$. Thus $\tilde{Q}_\mu\approx Q_\text{abs}$ for modes with $Q_\mu \gg Q_\text{abs}$, and $D^\mu_{th}\propto\tilde{Q}^{-1}_\mu$.

For a range of pump power $0<D_0<4D_\text{abs}$, we know from equation~(\ref{eq:Qopt}) that the optimally outcoupled mode has $Q_\mu \approx Q_\text{abs}$, or equivalently, $\tilde{Q}_\mu=Q_\text{abs}/2$.  Among the 56 modes within the full-width-at-half-maximum of the gain curve, there are two candidates close to this requirement: one with angular momentum $m=25$ and outcoupling $Q_\mu\simeq1.05\times10^4$, and another with $m=26$ and $Q_\mu\simeq2.18\times10^4$ [\Fig{fig:disk}(b)]. Since the former is closer to the gain center $\omega_aR=20.28$, it has a lower threshold [$2.58D_\text{abs}$ versus $2.88D_\text{abs}$; see \Fig{fig:disk}(a)] and we choose it as the target mode.

The spatial profile of the mode $m=25$ is shown in \Fig{fig:disk}(c). Like other whispering-gallery modes, it displays a maximum intensity on the ``caustic", a circle of radius $R_c \approx R \sin \chi_\mu$ \cite{JensThesis}. Therefore, we choose a ring-shaped pump profile/contact geometry shown in \Fig{fig:disk}(c). The numerically calculated threshold of this mode (obtained by solving the full set of non-linear equations (\ref{eq:SALT1})) is reduced by a factor of $5.52$ when compared with the threshold of the same mode in the uniformly pumped case [\Fig{fig:disk}(a)], agreeing with the $\re{f_\mu}$-factor in the pump focusing effect discussed in the previous section. For the higher-$Q$ modes, the caustic shifts gradually outward with $m$ and results in a reduced overlap with the selected pump profile. For example, the $m=31$ mode at $\Omega_\mu R=20.53$ exhibits an {\it increased} threshold by a factor of $5.31$, which is again well captured by $\re{f_\mu}^{-1}=5.38$. The thresholds of modes with even higher $m$ increase further, and they can be safely discarded when lasing is concerned.

We first discuss the full non-linear lasing solution with uniform pumping for $0<D_0<4D_\text{abs}$, shown in \Fig{fig:disk}(d) for the modes with the strongest output powers. At $D_0 = 4D_\text{abs}$ the output spectrum measured in the farfield is shown in \Fig{fig:disk}(e), upper panel, with ten modes lasing. The mode $m=28$ with $Q_\mu=5.66\times10^5$ has the strongest output and the optimally outcoupled mode $m=25$ is fully suppressed
by non-linear spatial hole-burning interactions,
regardless of how hard we pump.
Such insight is very hard to gain with Maxwell-Bloch simulations. In the formalism discussed in the SI, we have direct access to the exact evolution of modal gain for modes even below their lasing thresholds, shown in \Fig{fig:disk}(f). It reveals that the modal gain experienced by mode $m=25$
is strongly non-linear and saturates, as more and more modes turn on.

For the proposed ring-shaped pump, the laser remains single-mode up to $D_0 = 4D_\text{abs}$. Remarkably, the total output power contained in this single mode is 185 times that of the {\it total power} delivered by the ten lasing modes with uniform pumping. Considering that the total input power is identical, just distributed differently in the plane of the cavity, this implies that mode competition via spatial hole burning interactions can reduce the power efficiency of lasers substantially. We find this to be a very general statement valid for any laser cavity or gain mechanism as discussed in Section 2 of the SI.

\vspace{8pt}
\noindent\textbf{Discussion and conclusion} -- The enhancement of total output power with selective pumping proposed in this paper is universal. It derives from three effects: 1) much improved outcoupling due to the particular mode selected, 2) threshold reduction due to ``pump focusing", and 3) reduced gain competition with fewer modes. There will be more total output power at a given input power regardless of the cavity shape and gain mechanism, if the pumping distribution is chosen appropriately. When complemented with active-region quantum design \cite{e1,e2,e4,e5}, spatially selective pumping can provide devices operating at optimal operation conditions and deliver maximal power under given fabrication and pump power constraints.

The precise value of the power enhancement will depend on the particular cavity geometry and pump power. The cases of a stadium-shaped resonator, a photonic crystal defect-mode cavity, and a ridge-cavity microlaser are discussed in Section 2 of the SI, demonstrating the universality and robustness of our approach. Our general conclusion is that the largest enhancement factors can be expected at lower pump powers and for larger cavities.

As mentioned, our formalism and conclusions are also largely independent of the gain mechanism. We investigate this issue for a QCL gain medium as well as for an optically pumped 4-level gain medium in the \si, Section 2. In certain solid-state lasers, the presence of significant current-spreading can result in a sub-optimal distribution of current in the active region, reducing expected output power enhancement factors. This can be alleviated by engineering the contacts through which the current is injected such that spreading is taken into account. We investigate the robustness of our results to current spreading in the \si, Section 3.

Finally, we note that our conclusion implies that the deliberate introduction of loss in selected areas of a laser ultimately can increase the power efficiency of the device. This interesting fact implies that optical loss, which is traditionally avoided, can be a powerful tool when judiciously employed. The manipulation of the real part of the index of refraction of optical systems is today the cornerstone of modern photonics and quantum optics. The type of gain/loss engineering demonstrated here and its electrical (or optical) control promises a new and extended design space the full potential of which is to be explored in the coming years.

\section*{Acknowledgements}

We thank Nyan Aung, Hui Cao, Yamac Dikmelik, Dario Gerace, Claire Gmachl, Loan Le and Igor Trofimov for stimulating discussions. This work is supported by MIRTHE NSF EEC-0540832 and DARPA Grant No. N66001-11-1-4162.

\bibliographystyle{apsrev}

\end{document}